# COMPARISON OF SUB-GRID SCALE MODELS FOR LARGE-EDDY SIMULATION USING A HIGH-ORDER SPECTRAL ELEMENT APPROXIMATION OF THE COMPRESSIBLE NAVIER-STOKES EQUATIONS AT LOW MACH NUMBER

**Sohail Reddy**
Department of Applied Mathematics
Naval Postgraduate School
Monterey, CA
sredd001@fiu.edu

**Yassine Tissaoui**
Department of Mechanical and Industrial Engineering
New Jersey Institute of Technology
Newark, NJ
yt277@njit.edu

**Felipe A. V. de Bragança Alves**
Department of Applied Mathematics
Naval Postgraduate School
Monterey, CA
felipe.alves.br@nps.edu

**Simone Marras**
Department of Mechanical and Industrial Engineering
New Jersey Institute of Technology
Newark, NJ
smarras@njit.edu

**Francis X. Giraldo**
Department of Applied Mathematics
Naval Postgraduate School
Monterey, CA
fxgirald@nps.edu

March 4, 2021

## ABSTRACT

This study aims to identify the properties, advantages, and drawbacks of some common (and some less common) sub-grid scale (SGS) models for large eddy simulation of low Mach compressible flows using high order spectral elements. The models investigated and compared are the classical constant coefficient Smagorinsky-Lilly; the model by Vreman, which gained popularity among CFD practitioners in industrial and aerospace applications due to its minimal dissipation properties; and two flavors of a dynamic SGS (DSGS) model designed to stabilize finite and spectral elements for transport dominated problems. In particular, we compare one flavor of DSGS that is based on a time-dependent residual based version (R-DSGS) in contrast to a time-independent residual based scheme (T-DSGS). The SGS models are compared against the reference model by Smagorinsky and Lilly with respect to the ability to: (i) stabilize the numerical solution, (ii) minimize undershoots and overshoots, (iii) capture/preserve discontinuities, and (iv) transfer energy across different length scales. These abilities are investigated on problems for: (1) passively advected tracers, (2) coupled, nonlinear system of equations exhibiting discontinuities, (3) gravity-driven flows in a stratified atmosphere, and (4) homogenous, isotropic turbulence. All models show the ability to preserve sharp discontinuities. Vreman and the residual based DSGS models also reduce the undershoots and overshoots in the solution of linear and non-linear advection with sharp gradients. Our analysis shows that the R-DSGS and T-DSGS models are more robust than Vreman and Smagorinsky-Lilly when numerical stability of high-order spectral elements is the important metric of comparison. The Smagorinsky and Vreman models are better able to resolve the finer flow structures in shear flows, while the nodal R-DSGS model shows better energy conservation. Overall, the nodal implementation of R-DSGS (in contrast to its element-based counterpart) is shown to perform better than the other SGS models with respect to most of the metrics listed above, and on par with respect to the remaining ones.

# 1 Introduction

The seminal work by Smagorinsky [1] and Lilly [2] introduced the concept of Large-Eddy Simulations (LES), which has, ever since, developed into a widespread approach for turbulence research across fields [3,4]. LES is constructed by separating the grid-resolved large flow scales (large eddies) from the sub-grid scales, which are modeled by means of a subgrid-scale (SGS) model [5,6]. Since the introduction of the Smagorinsky-Lilly model (SL), several SGS turbulence closures have been proposed to overcome its limitations in the proximity of solid walls (e.g. [7–19].) Moreover, much effort has been dedicated to the comparison and identification of the benefits of various SGS models. Many have been analyzed using finite volume (FVM) [20–22], finite difference (FDM) [23,24], discontinuous Galerkin (DG) [25] and spectral element methods (SEM) [26, 27]. In the case of spectral and finite elements, some SGS models were not originally designed with LES in mind but, rather, as a way to stabilize the continuous Galerkin numerical solution of advection dominated problems (e.g. [28–32]). As a by-product, they also proved to be effective for turbulence modeling. This was seen in the case of DSGS, whose alternative implementations are assessed in this study, and the variational multi-scale stabilization (VMS) method by [33] as a turbulence modeling tool for LES using DG in, e.g., [34–37].

When used with element-based Galerkin methods (e.g., [38]), the SGS models that have appeared in the literature (e.g., [1, 9, 29]) assume a constant eddy diffusion within each element even when high-order elements are used (e.g. [39, 40]). In the current paper, we remove this constraint and allow the eddy diffusion to be variable within each high-order element. Besides testing these SGS models on standard turbulent test cases (as stabilization methods for the compressible Euler equations), we also test these methods as a means to stabilize tracer transport problems embedded within turbulent flows modeled by the compressible equations. Recently, work has appeared in the literature whereby the effective viscosity of high-order methods are studied [41] in, what is known as, implicit LES (or ILES) without the need to introduce additional SGS models [42]. This is very different (and quite intriguing) from the approach we pursue in the current work and only compare against standard SGS modeling approaches.

Fureby et al. [20] compared eight different SGS models on decaying homogenous isotropic turbulence. They showed that the choice of the SGS model is not critical for correctly modeling macroscopic flows if the spatial resolution is adequate. Rafei et al. [21] compared the classical Smagorinsky model with ILES on the Taylor Green vortex problem. Their work showed a non-physical increase in kinetic energy dissipation when the flow becomes disorganized on coarse meshes. This non-physical behavior disappeared for fine meshes. They also found that schemes with a numerical dissipation (due to the scheme) greater than the physical dissipation (due to the SGS model) can destroy the accuracy of the solution. Therefore, a comparison using spectral element discretization is ideal due to the lack of inherent numerical dissipation present. Tang et al. [23] compared the Smagorinsky, Vreman and a one-equation eddy viscosity model in an asymmetric diffuser geometry using a finite difference solution of the governing equations. They showed that the results of the one-equation and Vreman model agreed more closely with experimental and DNS results than the Smagorinsky-type models. Ghaisas et al. [22] also investigated the Smagorinsky, Vreman and the Sigma models on buoyant turbulent flows in a thermally heated cavity. They showed that the Smagorinsky and Sigma models showed good comparison to the DNS data, and stated that the Vreman model requires additional studies for buoyant turbulent flows. Anderson et al. [24] compared the less popular Wong-Lilly [43] SGS model and the Smagorinsky model for atmospheric boundary layer flows. They showed that the Wong-Lilly model is more dissipative than Smagorinsky's, which is known for being overly dissipative in the proximity of solid boundaries.

This study investigates the ability of various SGS models to resolve the flow, stabilize the numerical solution, and limit the amount of undershoots and overshoots while preserving the sharpness of the solution in the proximity of strong gradients. Four different SGS models and their variants are investigated on four test problems. These test problems range from canonical problems which admit analytical solution to problems with nonlinear, discontinuous solutions. They include passive advection of a tracer in deforming flow fields, two-dimensional Burgers' systems exhibiting discontinuities, homogenous, isotropic turbulence decay, and density current in atmospheric flows. The results are compared with those available in the literature.

The governing equations are described in § 2. The SGS models are described in § 3. The construction of the spectral element method is detailed in § 4. The different test cases are described in § 5 through § 8, which are followed by concluding remarks in § 9.

## 2 Governing Equations

This study considers flow governed by the time-dependent compressible Euler equations. Let $\Omega \in \mathbb{R}^3$ be a fixed domain in 3D with boundary $\Gamma$. Let us denote the fluid density, velocity vector, potential temperature and the tracer concentration by $\rho$, $\mathbf{u} = u_i$ $(i = 1, 2, 3)$, $\theta$ and $q$. Then, the time-dependent equations of mass, momentum, potential temperature (used in lieu of the energy equation for reasons summarized in, e.g., [44]), and tracer transport equations in non-flux form can be written as:

$$\frac{\partial \rho}{\partial t} + \frac{\partial \rho u_j}{\partial x_j} = 0 \tag{1a}$$

$$\frac{\partial u_i}{\partial t} + u_j \frac{\partial u_i}{\partial x_j} + \frac{1}{\rho} \frac{\partial p}{\partial x_i} = -g \delta_{i3} \tag{1b}$$

$$\frac{\partial \theta}{\partial t} + u_j \frac{\partial \theta}{\partial x_j} = 0 \tag{1c}$$

$$\frac{\partial q}{\partial t} + u_j \frac{\partial q}{\partial x_j} = 0 \tag{1d}$$

where summation is implied for repeated indices, $g$ is the acceleration due to gravity, $\delta_{ij}$ is the kronecker delta, and the pressure, $p$, is related to $\theta$, and $\rho$ through the equation of state for a perfect gas

$$p = p_0 \left( \frac{\rho R \theta}{p_0} \right)^{\gamma}.$$

Eqs. (1) must be solved in $\Omega \, \forall \, t \in \mathbb{R}^+$ given proper initial and boundary conditions. As is often the case in Numerical Weather Prediction (NWP), the thermodynamic quantities $(\rho, \theta, p)$ are split into a hydrostatically balanced reference state $(\rho_R, \theta_R, p_R)$ and perturbation $(\Delta \rho, \Delta \theta, \Delta p)$ as $\Delta \rho(t, \mathbf{x}) = \rho(t, \mathbf{x}) - \rho_R(z)$, $\Delta \theta(t, \mathbf{x}) = \theta(t, \mathbf{x}) - \theta_R(z)$, and $\Delta p(t, \mathbf{x}) = p(t, \mathbf{x}) - p_R(z)$. The system of equations (1), referred to as 'Set2NC' in [45], are solved within the Nonhydrostatic Unified Model of the Atmosphere (NUMA) [46, 47].

### 2.1 Large-Eddy Simulation (LES)

The LES formulation of the problem above can be obtained by first introducing the spatial filtering operation

$$\overline{f}(\mathbf{x}) = \int_{\Omega} G_{\overline{\Delta}}(\mathbf{x} - \boldsymbol{\chi}) f(\boldsymbol{\chi}) \, d\boldsymbol{\chi} \tag{2}$$

where the grid resolved quantity, $\overline{f}$, is obtained from the filtering function $G$ of the instantaneous quantities, $f$, using a filter width $\overline{\Delta}$. The application of Eq. (2) to the continuity equation (Eq. (1a)) results in the presence of an additional sub-grid term on the right-hand side. This can be avoided using the density-weighted filtering called Favre filtering $\widetilde{\phi} = \overline{\rho \phi} / \overline{\rho}$ [48][1]. The two operations yield the filtered equations

$$\frac{\partial \overline{\rho}}{\partial t} + \frac{\partial \overline{\rho} \widetilde{u}_j}{\partial x_j} = 0, \tag{3a}$$

$$\frac{\partial \widetilde{u}_i}{\partial t} + \widetilde{u}_j \frac{\partial \widetilde{u}_i}{\partial x_j} + \frac{1}{\overline{\rho}} \frac{\partial \overline{p}}{\partial x_i} = -\frac{1}{\overline{\rho}} \frac{\partial \tau_{ij}^{SGS}}{\partial x_j} - g \delta_{i3}, \tag{3b}$$

$$\frac{\partial \widetilde{\theta}}{\partial t} + \widetilde{u}_j \frac{\partial \widetilde{\theta}}{\partial x_j} = -\frac{1}{\overline{\rho}} \frac{\partial \Theta_j^{SGS}}{\partial x_j}, \tag{3c}$$

$$\frac{\partial \widetilde{q}}{\partial t} + \widetilde{u}_j \frac{\partial \widetilde{q}}{\partial x_j} = -\frac{\partial Q_j^{SGS}}{\partial x_j} \tag{3d}$$

where the derivatives of $\tau_{ij}^{SGS}$, $\Theta_j^{SGS}$ and $Q_j^{SGS}$ on the right-hand side of Eq. (3b), Eq. (3c) and Eq. (3d) represent the contribution of the unresolved sub-grid scales (SGS). In Eq. (3b), $\tau_{ij}^{SGS}$ is the turbulent stress tensor,

$$\tau_{ij}^{SGS} = \overline{\rho} \left( \widetilde{u_i u_j} - \widetilde{u}_i \widetilde{u}_j \right)$$

---

[1]From now on, the symbols $\overline{\cdot}$ and $\widetilde{\cdot}$ will indicate the grid-resolved quantities (large-eddy).

which is related to unresolved scales and must, therefore, be modeled. Our study focuses on eddy viscosity models, where the turbulent stress tensor is approximated as a function of the strain rate tensor

$$S_{ij} = \frac{1}{2} \left( \frac{\partial \widetilde{u}_i}{\partial x_j} + \frac{\partial \widetilde{u}_j}{\partial x_i} \right)$$

as

$$\tau_{ij}^{SGS} = -2\mu S_{ij}. \tag{4}$$

The coefficient $\mu$ is constructed using different SGS models. Similarly, the quantity $\Theta_j^{SGS}$ in Eq. (3c) resulting from filtering Eq. (1c) is given by

$$\Theta_j^{SGS} = \overline{\rho} \left( \widetilde{\theta u_j} - \widetilde{\theta} \widetilde{u}_j \right) \tag{5}$$

and is modeled as

$$\Theta_j^{SGS} = -\kappa \frac{\partial \widetilde{\theta}}{\partial x_j} \tag{6}$$

where the coefficient $\kappa$ is defined in Sec. 3. The quantity $Q_j^{SGS}$ is also expressed similarly as

$$Q_j^{SGS} = -\alpha \frac{\partial \widetilde{q}}{\partial x_j} \tag{7}$$

where the construction of the coefficient $\alpha$ is defined in Sec. 3. This work uses implicit filtering where the grid is assumed to be the LES low-pass filter. Therefore, the values of the prognostic variables at the quadrature points are taken to be the filtered quantities.

## 3   Sub-grid Scale Models

This study considers four sub-grid scale models: Smagorinsky-Lilly, Vreman, R-DSGS, and T-DSGS. The formulation of each model is described below.

### 3.1   Smagorinsky-Lilly Model

The Smagorinsky-Lilly turbulence eddy viscosity model [1,49] defines the diffusion coefficient as a quadratic function of a filter-width $\Delta$ and the magnitude of the rate of strain tensor $S_{ij} = (\partial_j u_i + \partial_i u_j)$. The filter width in this study coincides with the effective grid resolution in an isotropic grid. The definition of the filter width used in this work is given in Sec. 3.4. For an anisotropic grid, the filter width can be selected according to the schemes suggested in Table 1. The Smagorinsky-Lilly model, here onwards referred to as the Smagorinsky model, or abbreviated as SL, defines the turbulent eddy viscosity as

$$\nu_t = (C_s \Delta)^2 \sqrt{2S_{ij}^2}, \tag{8}$$

where $C_s$ is a constant coefficient usually taken to be in the range $0.12 < C_s < 0.21$. For all cases in this work, $C_s$ is held constant at 0.14 [39].

### 3.2   Vreman Model

The Vreman SGS model [9] is a popular choice by CFD modelers for industrial and aerospace applications because of its robustness across flow regimes and because it has low dissipation near wall boundaries and in transitional flow. Its computational complexity is similar to the classical SL model [1, 2]. To the authors' knowledge, Vreman has not been previously studied within the spectral element framework or in the context of atmospheric flows.

The turbulent eddy viscosity of this model depends on the first derivatives of the velocity vector components and is given by

$$\nu_t = c \sqrt{\frac{B_\beta}{\frac{\partial u_i}{\partial x_j} \frac{\partial u_i}{\partial x_j}}}, \tag{9a}$$

where $c \approx 2.5 C_s^2$ and

$$B_\beta = \beta_{11}\beta_{22} + \beta_{11}\beta_{33} + \beta_{22}\beta_{33} - (\beta_{13}^2 + \beta_{12}^2 + \beta_{23}^2),$$
$$\beta_{ij} = \Delta_k^2 \frac{\partial u_i}{\partial x_k} \frac{\partial u_j}{\partial x_k} . \tag{9b}$$

It should be noted that $\dfrac{\partial u_i}{\partial x_j}\dfrac{\partial u_i}{\partial x_j}$ and $B_\beta$ are the first and second principle invariants, respectively, of the velocity vector gradient. The mixing lengths $\Delta_k$ can be determined as the effective resolution along the direction $k$. Due to the isotropy of grids used in this work, the effective resolution is the same in all directions.

### 3.3 Dynamic Sub-Grid Scale (DSGS) Model

The residual-based Dynamic Sub-Grid Scale (DSGS) model [29] is a relatively new approach for constructing the eddy viscosity. Originally designed to stabilize high order continuous and discontinuous spectral element solvers [39, 50] and capture sharp gradients [51], it builds $\mu$, $\kappa$ and $\alpha$ using the residuals of the filtered equations.

Consider the system given by Eq. (3) where the dissipative terms are excluded. The residuals are then defined as

$$R(\overline{\rho}) = \frac{\partial \overline{\rho}}{\partial t} + \frac{\partial \overline{\rho}\,\widetilde{u}_j}{\partial x_j}, \tag{10a}$$

$$R(\widetilde{u}_i) = \frac{\partial \widetilde{u}_i}{\partial t} + \widetilde{u}_j \frac{\partial \widetilde{u}_i}{\partial x_j} + \frac{1}{\overline{\rho}}\frac{\partial \overline{p}}{\partial x_i} + g\delta_{i3}, \tag{10b}$$

$$R(\widetilde{\theta}) = \frac{\partial \widetilde{\theta}}{\partial t} + \widetilde{u}_j \frac{\partial \widetilde{\theta}}{\partial x_j}, \tag{10c}$$

$$R(\widetilde{q}) = \frac{\partial \widetilde{q}}{\partial t} + \widetilde{u}_j \frac{\partial \widetilde{q}}{\partial x_j}. \tag{10d}$$

The time derivatives in Eq. (10) are approximated using a first-order backward difference formula. No significant differences were observed between the first order and second order finite difference approximation for the time derivative. To define the diffusion coefficients we first define its maximum allowable value as

$$\mu_{\max} = 0.5\overline{\Delta}\left\|\,|\widetilde{\mathbf{u}}| + \sqrt{\gamma \frac{\overline{p}}{\overline{\rho}}}\,\right\|_\infty, \tag{11}$$

where $|\widetilde{\mathbf{u}}| + \sqrt{\gamma \overline{p}/\overline{\rho}}$ is the maximum wave speed. We then define the residual-based diffusion coefficient $\mu_{res}$ as

$$\mu_{res} = C\overline{\Delta}^2 \max\left(\frac{\|R\left(\widetilde{\mathbf{u}}\right)\|_\infty}{\|\widetilde{\mathbf{u}} - \widehat{\mathbf{u}}\|_{\infty,\Omega}}\right) \tag{12}$$

where $C$ is a scaling parameter similar to $C_s$ in the Smagorinsky model, $\widehat{\cdot}$ is the space averaged quantity over $\Omega$ and $\|\cdot\|_{\infty,\Omega}$ are the normalization constants used to preserve the correct dimension of the resulting equations. Similarly, $\kappa_{res}$ and $\alpha_{res}$ are defined as

$$\kappa_{res} = C\overline{\Delta}^2\left(\frac{\|R(\widetilde{\theta})\|_\infty}{\|\widetilde{\theta} - \widehat{\theta}\|_{\infty,\Omega}}\right), \tag{13a}$$

$$\alpha_{res} = C\overline{\Delta}^2\left(\frac{\|R(\widetilde{q})\|_\infty}{\|\widetilde{q} - \widehat{q}\|_{\infty,\Omega}}\right). \tag{13b}$$

We set $C = 1$ for all DSGS models in this work. The diffusion coefficients are defined as

$$\mu = \min\left(\mu_{\max}, \mu_{res}\right), \tag{14a}$$

$$\kappa = \frac{\text{Pr}}{\gamma - 1}\min\left(\mu_{\max}, \kappa_{res}\right), \tag{14b}$$

$$\alpha = \frac{\text{Pr}}{\gamma - 1}\min\left(\mu_{\max}, \alpha_{res}\right), \tag{14c}$$

where Pr=0.7 is an artificial Prandtl number. It can be seen that the dissipation is proportional to the residual of the governing equation which is proportional to $\overline{\Delta}_{\Omega_e}^{-1}$ at discontinuities and is relatively small near smooth regions. The definition of dynamic viscosity given by Eq. (14a) restricts the maximum to $\mu_{\max}$ which is equivalent to a stable upwind scheme in regions with sharp discontinuities. In the remainder of this work, this approach for constructing the diffusion coefficients based on residuals will be referred to as residual-based dynamic sub-grid scale (R-DSGS) model. This study also considers an alternative approach for constructing the diffusion coefficients where all of the temporal terms are omitted from Eq. (10). That is, the construction of this model is similar to the R-DSGS except it does not include the temporal derivatives in the residual terms in Eq. (10). Therefore, this model will be referred to as transient-based dynamic sub-grid scale (T-DSGS) model. The diffusion coefficients obtained using this model are proportional to the transience ($\partial/\partial t$) of the flow.

It should be mentioned that two different representations for the diffusion coefficients are used. The first approach, denoted by the suffix 'E' (for elemental), assumes that the coefficient is uniform over an element and piecewise uniform over the domain. The second, denoted by the suffix 'N' (for nodal), represents the diffusion coefficient by the same order basis expansion as the prognostic variables. The elementally uniform definition of the diffusion coefficient is obtained by replacing $\|\cdot\|_\infty$ in Eq. (11), Eq. (12), and Eq. (13a) by $\|\cdot\|_{\infty,\Omega_e}$. That is, the infinity norm is taken over the set of all LGL points over an element $\Omega_e$. For the nodal representation, the infinity norm is simply the absolute value of the residual at the LGL points. The elemental formulations are expected to be more dissipative since the diffusion coefficients for the element are taken as the maximum over the set of LGL points of the element (i.e., infinity norm). Therefore, a nodal approach is expected to yield more resolved flows since the diffusion coefficient depends on the properties at only that LGL point and not the entire set.

### 3.4 Selection of Filter Width

Each SGS model requires the definition of a filter width $\Delta$. The computational meshes used in this work are isotropic, where the edge lengths of elements are of relatively equal lengths. Then, for an element $\Omega_e$ of order $N$ with edge lengths $\Delta x, \Delta y, \Delta z$, the characteristic length and, hence, filter width is defined as

$$\overline{\Delta} = \min\left(\Delta x, \Delta y, \Delta z\right)/(N+1).$$

This definition is sufficient given the isotropic nature of the mesh in the current study. A more proper definition of $\overline{\Delta}$ for LES should be used for highly anisotropic meshes. Such mesh anisotropy is seen in global climate models where the horizontal dimension of the globe is orders of magnitude greater than the vertical dimension. For such cases, the filter width should be appropriately chosen. Although not an exhaustive list, a few possible definitions of the filter width are given in Table 1.

Table 1: Definitions of element characteristic length for anisotropic grids.

| Formula | Notes | References |
|---|---|---|
| $\overline{\Delta} = \left(\Delta x \Delta y \Delta z\right)^{1/3}/N$ | | [52, 53] |
| $\overline{\Delta} = f\left(\Delta x \Delta y \Delta z\right)^{1/3}/N$ | | [54] |
| where | $f = \cosh\left[\sqrt{\frac{4}{27}\left(\ln^2 a_1 + \ln^2 a_2 - \ln a_1 \ln a_2\right)}\right]^{1/2}$ | |
| | $a_{1,2} = \Delta_{1,2}/\max\left(\Delta x, \Delta y, \Delta z\right)/(N+1)$ | |
| | $\Delta_{1,2}$ are the two smallest edge length | |
| | and $N$ is the order of approximation | |

## 4 Spectral Element Discretization

The governing equations (1) are solved via a continuous spectral element discretization in space. Consider a system of partial differential equations (PDEs) in compact vector notation

$$\frac{\partial \mathbf{q}}{\partial t} + \nabla \cdot \mathcal{F}\left(\mathbf{q}\right) = \mathcal{S}\left(\mathbf{q}\right) \tag{15}$$

where $\mathbf{q} = \left\{\overline{\rho}, \widetilde{\mathbf{u}}, \widetilde{\theta}, \widetilde{q}\right\}$. Multiplying Eq. (15) by a test function $\psi$ and integrating over the domain yields

$$\int_\Omega \psi\left(\mathbf{x}\right)\frac{\partial \mathbf{q}}{\partial t}d\Omega + \int_\Omega \psi\left(\mathbf{x}\right)\nabla \cdot \mathcal{F}\left(\mathbf{q}\right)d\Omega = \int_\Omega \psi\left(\mathbf{x}\right)\mathcal{S}\left(\mathbf{q}\right)d\Omega. \tag{16}$$

The second term can be expanded via the product rule ($\psi\nabla \cdot \mathcal{F} = \nabla \cdot \left(\psi\mathcal{F}\right) - \nabla\psi \cdot \mathcal{F}$). Using an element-wise formulation, Eq. (16) then becomes

$$\int_{\Omega_e} \psi_e\frac{\partial \mathbf{q}}{\partial t}d\Omega_e + \int_{\Gamma_e} \psi_e\hat{\mathbf{n}} \cdot \mathcal{F}\left(\mathbf{q}\right)d\Gamma_e - \int_{\Omega_e} \nabla\psi \cdot \mathcal{F}\left(\mathbf{q}\right)d\Omega_e = \int_{\Omega_e} \psi_e\mathcal{S}\left(\mathbf{q}\right)d\Omega_e \tag{17}$$

where we have invoked the divergence theorem in the second term. For the diffusion operators, we begin with an ad hoc viscous operator

$$\nabla \cdot (\mu \nabla \mathbf{q})$$

where $\mu = \mu(\mathbf{x})$, as will be the case for nodally defined diffusion coefficients. Multiplication by a test function $\psi$ and integration over the domain yields

$$\int_{\Omega} \psi(\mathbf{x}) \nabla \cdot (\mu \nabla \mathbf{q}) d\Omega.$$

Using integration by parts, the element-wise formulation can be written as

$$\int_{\Omega_e} \psi_e(\mathbf{x}) \nabla \cdot (\mu \nabla \mathbf{q}) d\Omega_e = \int_{\Gamma_e} \mu \psi_e (\hat{\mathbf{n}_e} \cdot \nabla \mathbf{q}) \ d\Gamma_e - \int_{\Omega_e} \mu \nabla \psi_e \cdot \nabla \mathbf{q} \ d\Omega_e. \tag{18}$$

We can represent each variable in $\mathbf{q}$ as a linear combination of basis functions

$$\mathbf{q}^e(\mathbf{x}, t) = \sum_{j=1}^{(N+1)^3} \mathbf{q}_j^e(t) \, \psi_j(\mathbf{x}) \tag{19}$$

where the 3D basis functions are formed as the tensor product of 1D Lagrange polynomials ($\psi_{ijk}(\xi, \eta, \zeta) = \psi_i(\xi) \otimes \psi_j(\eta) \otimes \psi_k(\zeta)$) given by

$$\psi_i(\chi) = \prod_{j=1, j \neq i}^{N+1} \frac{\chi - \chi_j}{\chi_i - \chi_j} \tag{20}$$

where $N + 1$ is the number of points in the $\chi$ direction. Here, the Lagrange polynomials $\psi_i(\chi)$ are associated with a specific set of points, chosen here to be the Legendre-Gauss-Lobatto (LGL) points $\{\chi_i\} \in [-1, 1]$, which are the roots of

$$\left(1 - \chi^2\right) \mathcal{P}'_N(\chi)$$

where $\mathcal{P}'_N(\chi)$ is the first derivative of the the $N^{th}$ degree Legendre polynomial. The LGL points are also used for integration with quadrature weights given by

$$\omega_i = \frac{2}{N(N+1)} \left(\frac{1}{\mathcal{P}_N(\chi_i)}\right)^2. \tag{21}$$

Since the interpolation points and integration points are collocated, then it can be seen from Eq. (20) that $\psi_i(\chi_j) = \delta_{ij}$, thereby yielding a diagonal mass matrix. It should be mentioned that, because the solution, represented by Eq. (19), is continuous across element interfaces, the flux terms (second term in Eq. (17) and first term on the right in Eq. (18)) need only be evaluated at the physical boundary and not the inter-element boundaries.

The final step in the continuous spectral element method requires invoking the *direct stiffness summation* (DSS) operation whereby the element-wise solutions comprised of the combination of Eq. (17) and Eq. (18) are summed globally in order to construct the global solution (see, e.g., [45]). Using DSS yields

$$\sum_{e=1}^{N_e} \left[ \int_{\Omega_e} \psi_e \frac{\partial \mathbf{q}}{\partial t} d\Omega_e + \int_{\Gamma_e} \psi_e \hat{\mathbf{n}} \cdot \mathcal{F}(\mathbf{q}) d\Gamma_e - \int_{\Omega_e} \nabla \psi \cdot \mathcal{F}(\mathbf{q}) d\Omega_e \right] \tag{22}$$

$$= \sum_{e=1}^{N_e} \left[ \int_{\Omega_e} \psi_e \mathcal{S}(\mathbf{q}) \, d\Omega_e + \int_{\Gamma_e} \mu \psi_e (\hat{\mathbf{n}_e} \cdot \nabla \mathbf{q}) \ d\Gamma_e - \int_{\Omega_e} \mu \nabla \psi_e \cdot \nabla \mathbf{q} \ d\Omega_e \right] \tag{23}$$

which, defines a map $\Psi : (e, i) \rightarrow I$ that allows us to store the contribution of all local degrees of freedom $i = 1, \ldots (N+1)^3$ of all elements $e = 1, \ldots, N_e$ and store them in the global degrees of freedom $I = 1, \ldots, N_p$.

## 5    Case 1: Tracers in Planar Deformational Flow

### 5.1    Problem Definition

One of the most common test cases in the field of geophysical fluid dynamics is the transport of passive tracers in a deformational velocity field. These cases often feature a flow reversal to return the tracer to its initial distribution, thereby admitting an analytical solution. This case has been adopted from the work of [55, 56]. The deformational velocity field for this case is given by

$$u(x, y, t) = u_0 \sin^2(\pi x) \sin(2\pi y) \cos(\pi t/T)$$
$$v(x, y, t) = -u_0 \sin^2(\pi y) \sin(2\pi x) \cos(\pi t/T)$$

(24)

where the period $T$ was set at 24 hrs and $u_0$ was taken to be $4/T$. The velocity profile given by Eq. (24) results in a flow reversal at $T/2$. Therefore, the initial ($t = 0$) and final ($t = 24$ hrs) tracer distributions should be identical. The initial tracer concentration distribution features two discontinuous slotted-cylinders given by

$$q = \begin{cases} q_{max} & r_i/r_0 \leq 1 \quad \text{and} \quad (|x - x_i| \geq r_0/6 \quad \text{or} \quad y < y_i) \quad i = 1, 2 \\ 0 & \text{otherwise} \end{cases}$$

(25)

where $r_i = \sqrt{(x - x_i)^2 + (y - y_i)^2}$, $r_0 = 0.25$ and $q_{max} = 1.0$. These slotted cylinders are centered at $(x_1, y_1) = (0.7\,\text{m}, 0.5\,\text{m})$ and $(x_2, y_2) = (1.3\,\text{m}, 0.5\,\text{m})$. This test allows us to assess the ability of the SGS models to preserve a sharp interface.

The domain of analysis was taken to be the rectangle $\Omega(x, y) \in [0, 2] \times [0, 1]\,\text{m}^2$ with doubly periodic boundary conditions. The solution to the transport equation (Eq. (1d)) was computed on a $80 \times 40$ element grid with fourth-order polynomials for an effective resolution of $\Delta x = \Delta y \approx 6.25 \times 10^{-3}$ m and was advanced in time until $t = 24$ hrs. Figure 1 shows the initial distribution and exact (analytical) solution for the tracer distribution after $t = 24$ hrs. The time integration was performed using an explicit third-order five-stage Runge-Kutta method [57].

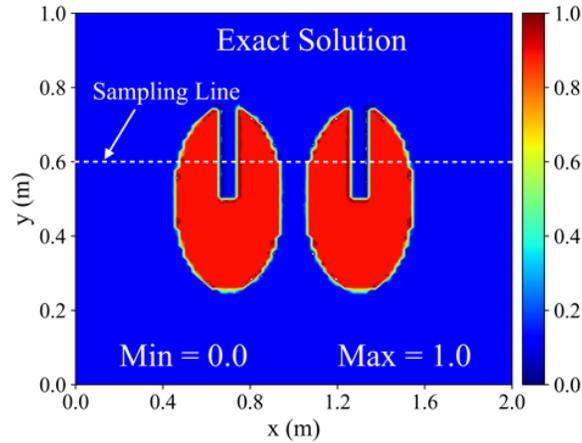

Figure 1: Initial and exact (analytical) solution for tracer distribution in 2D deformational swirling flow after 24 hours.

### 5.2    Results

This test case investigates the effects of SGS models on discontinuities and their ability to damp overshoots and undershoots while preserving the sharpness of the solution. The eddy viscosities for the Smagorinsky and Vreman models are obtained using the velocity profile given by Eq. (24) while the viscosity for the DSGS models is obtained using Eq. (14c), effectively decoupling the SGS model for the tracer from the velocity fields. Alternatively, Eq. (12) can be used thereby also taking into account the velocity field.

Figure 2 shows the tracer distribution after 24 hours using the different SGS models. As mentioned, the initial and final tracer distribution and extrema should be equal. Figure 2a shows the tracer distribution when using the Smagorinsky model. It can be seen that, although the model is able to preserve the sharp discontinuity, it is not able to preserve the extrema. This model results in an absolute error of 0.12 kg/kg and 0.15 kg/kg in the global minimum and maximum, respectively

The Vreman model Fig. (2b) is also able to preserve the discontinuity relatively well while also resulting in a lower error in the extrema than the Smagorinsky model. The absolute error in the global minimum and maximum is approximately 0.05 kg/kg. Figures 2c and 2e show the tracer concentration obtained using the elemental and nodal formulations of the R-DSGS model. The nodal formulation is better able to capture the discontinuous interface whereas the elemental formulation leads to a more diffused interface. This is due to the higher diffusion coefficients arising from the infinity-norm in the element-wise formulation. A similar behaviour is seen for the T-DSGS model (Fig. 2d and Fig. 2f). Both formulations, however, are able to limit overshoots and undershoots in the extrema. The elemental T-DSGS model (Fig. 2d) yields the most diffused interface and the smallest error in the global extrema. This is expected as the R-DSGS model is less diffusive since the magnitude of the residual is typically smaller than the unsteadiness of the flow. This increased dissipation of the elemental T-DSGS model also damps the Gibbs' phenomena (overshoots and undershoots). Figure 2 shows that the Vreman and nodal R-DSGS models offer a good balance between preserving the discontinuous interface and reducing the overshoots and undershoots.

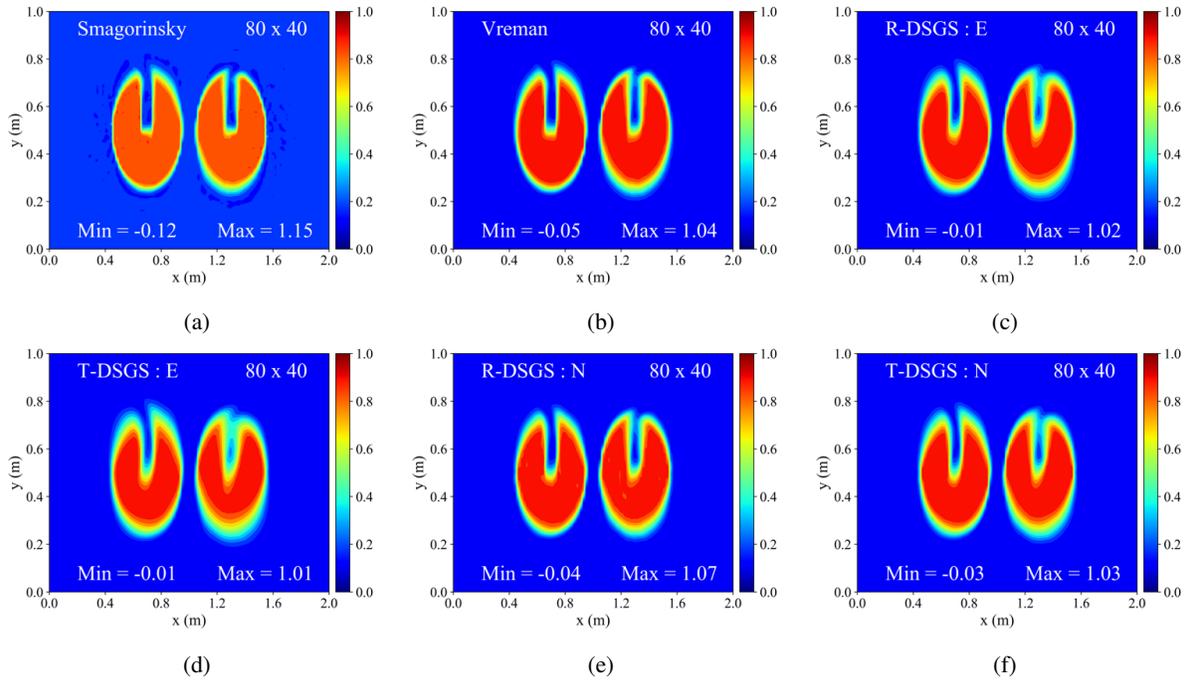

Figure 2: Tracer distribution after 24 hours for 2D deformational flow using: a) Smagorinsky, b) Vreman, c) elemental R-DSGS, d) elemental T-DSGS, e) nodal R-DSGS and f) nodal T-DSGS.

Figure 3 shows the tracer concentration along the sample line shown in Fig. 1. It shows that the Smagorinsky model is best able to preserve the discontinuity but results in larger overshoots and undershoots near the discontinuities. The remaining models yield a more diffused interface but also smaller overshoots and undershoots near the discontinuities.

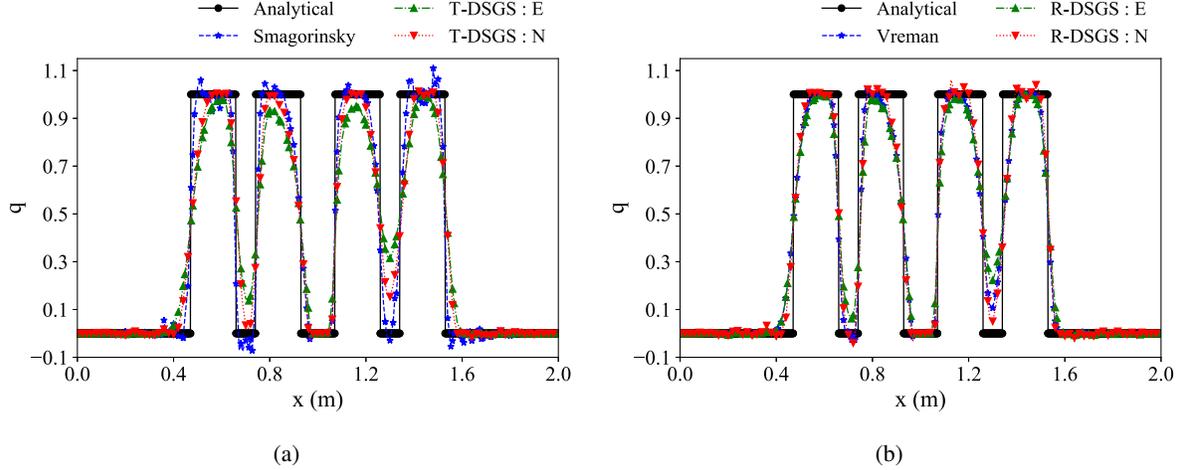

Figure 3: Tracer distribution after 24 hours for 2D deformational flow along a slice plane at $y = 0.6m$ using: a) Smagorinsky and T-DSGS models and b) Vreman and R-DSGS models.

## 6 Case 2: 2D Burgers' Equation

### 6.1 Problem Definition

The Burgers' equation has attracted much attention in the field of applied mathematics, fluid mechanics and acoustics. First introduced by Bateman [58] in 1915 and later studied by Burgers [59], the Burgers' equation has been used as a mathematical model of turbulence [60–62] and an approximate model of shock waves in viscous and inviscid fluids [63]. This problem considers a system of two non-linear conservation equations in 2D. Like the transport problem in Case 1, the inviscid Burgers' system also exhibits discontinuities but, unlike Case 1, it features a set of two non-linearly coupled equations. This case investigates the ability of the SGS model to not only resolve the discontinuity but also its ability to stabilize the non-linear solution. The Burgers' system is given by

$$\frac{\partial u}{\partial t} + u\frac{\partial u}{\partial x} + v\frac{\partial u}{\partial y} = 0 \tag{26a}$$

$$\frac{\partial v}{\partial t} + u\frac{\partial v}{\partial x} + v\frac{\partial v}{\partial y} = 0 \tag{26b}$$

which is identical to an isobaric form of Eq. (1b) without the gravitational body force. The initial conditions are given by $u(x, y, t = 0) = x + y$ and $v(x, y, t = 0) = x - y$. The domain is taken to be a unit square $\Omega(x, y) \in [0, 1]\mathrm{m} \times [0, 1]\mathrm{m}$ with doubly periodic boundary conditions [64]. The solution is advanced in time until $t = 1$ s and computed on a $60 \times 60$ element grid with fourth-order polynomials for an effective resolution of $\Delta x \approx \Delta y \approx 4.2 \times 10^{-3}$ m. The time integration is performed using an explicit third-order five-stage Runge-Kutta method. The value of the $C_s$ constant in the Smagorinsky and Vreman models is increased to 0.21 in an effort to better stabilize the nonlinear set of Burgers' equations. It should be mentioned that the DSGS models are used to compute two separate diffusion coefficients; one for each of the equations in Eq. (26). This is done by considering each velocity residual in Eq. (12) individually.

### 6.2 Results

The velocity magnitude ($\|\vec{V}\| = \sqrt{u^2 + v^2}$) at $t = 0.7$ s using the six SGS models is shown in Fig. 4. The discontinuous interfaces are better preserved (sharper) when using the Smagorinsky and Vreman models and are slightly more diffused when using the DSGS models. This is simply because the Smagorinsky and Vreman models result in smaller diffusion coefficients. It should be mentioned that the diffusion coefficients of the DSGS models can also be scaled by selecting an appropriate value of the parameter $C$ in Eq. (12) and Eq. (13). The Smagorinsky and Vreman SGS models result in oscillations in the solution near the discontinuities. These oscillations are seen in $x, y \in [0.6, 0.8]$. It should be noted that the oscillations are also present in regions without sharp discontinuities. The DSGS models (both R-DSGS and T-DSGS), however, damp these oscillations while preserving the sharp discontinuities.

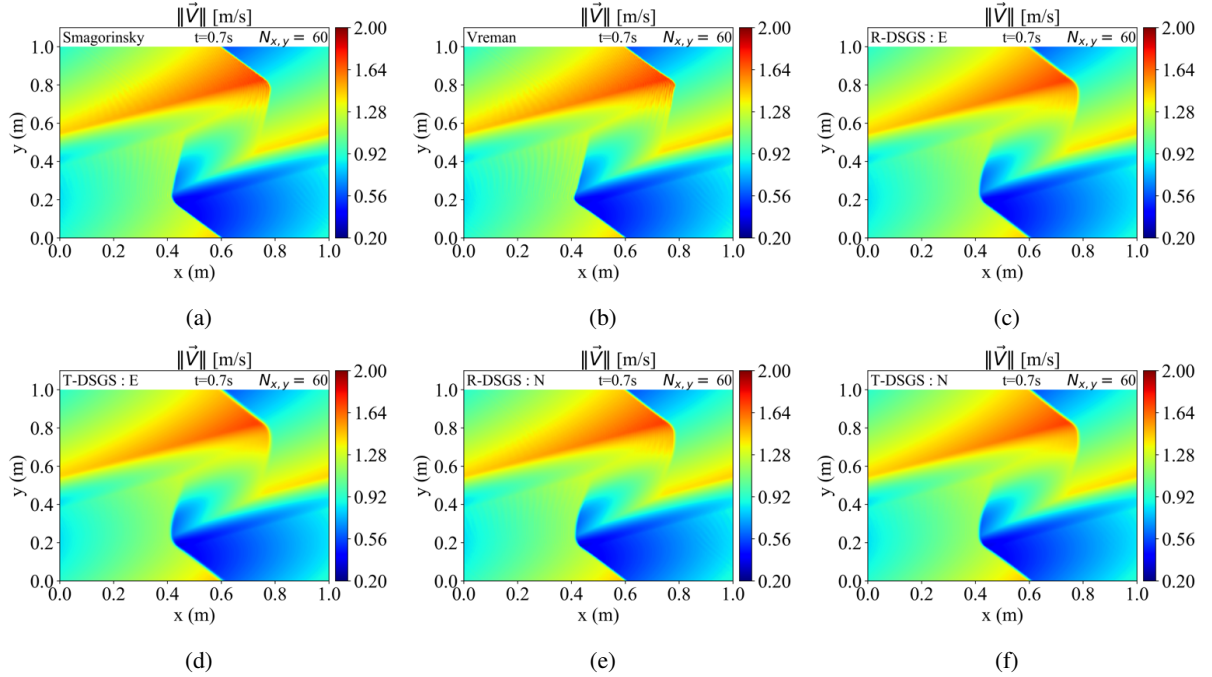

Figure 4: The velocity magnitude for the solution of the 2D Burgers' equation at time $t = 0.7\ s$ using: a) Smagorinsky, b) Vreman, c) elemental R-DSGS, d) elemental T-DSGS, e) nodal R-DSGS and f) nodal T-DSGS.

Figure 5 shows the $x$-velocity ($u$), $y$-velocity ($v$) and the distribution of the diffusion coefficients at $t = 1$ s. The solution using the Vreman model diverges at $t = 0.7$ s due to the unbounded growth of the oscillations near the discontinuities, thereby driving the solution unstable. This is because the diffusion coefficient computed by the model is used to stabilize both equations in Eq. (26). The DSGS models overcome this by decoupling the two equations and stabilizing each equation individually using different diffusion coefficients. The stabilization of the DSGS models damps these oscillations and stabilizes the solution.

It can be seen that although the Smagorinsky model does not diverge, it does result in significant oscillations near the discontinuities. Comparing R-DSGS and T-DSGS models to the Smagorinsky model shows that the DSGS models do not exhibit such oscillations near the discontinuous interfaces. Figures 5h and 5i show the distribution of the diffusion coefficients for Eq. (26a) and Eq. (26b), respectively. It shows that the distribution of the diffusion coefficients coincides well with the discontinuities in Figs. 5b and 5e and diminish further away from the discontinuities. The distribution obtained using the Smagorinsky model (Fig. 5g) follows the distribution of the modulus of the strain rate tensor and, therefore, the combined effects of both equations in the Burgers' system. The values of $\mu_x$ and $\mu_y$ for the T-DSGS are similar in their distributions but larger in magnitudes than those obtained using R-DSGS.

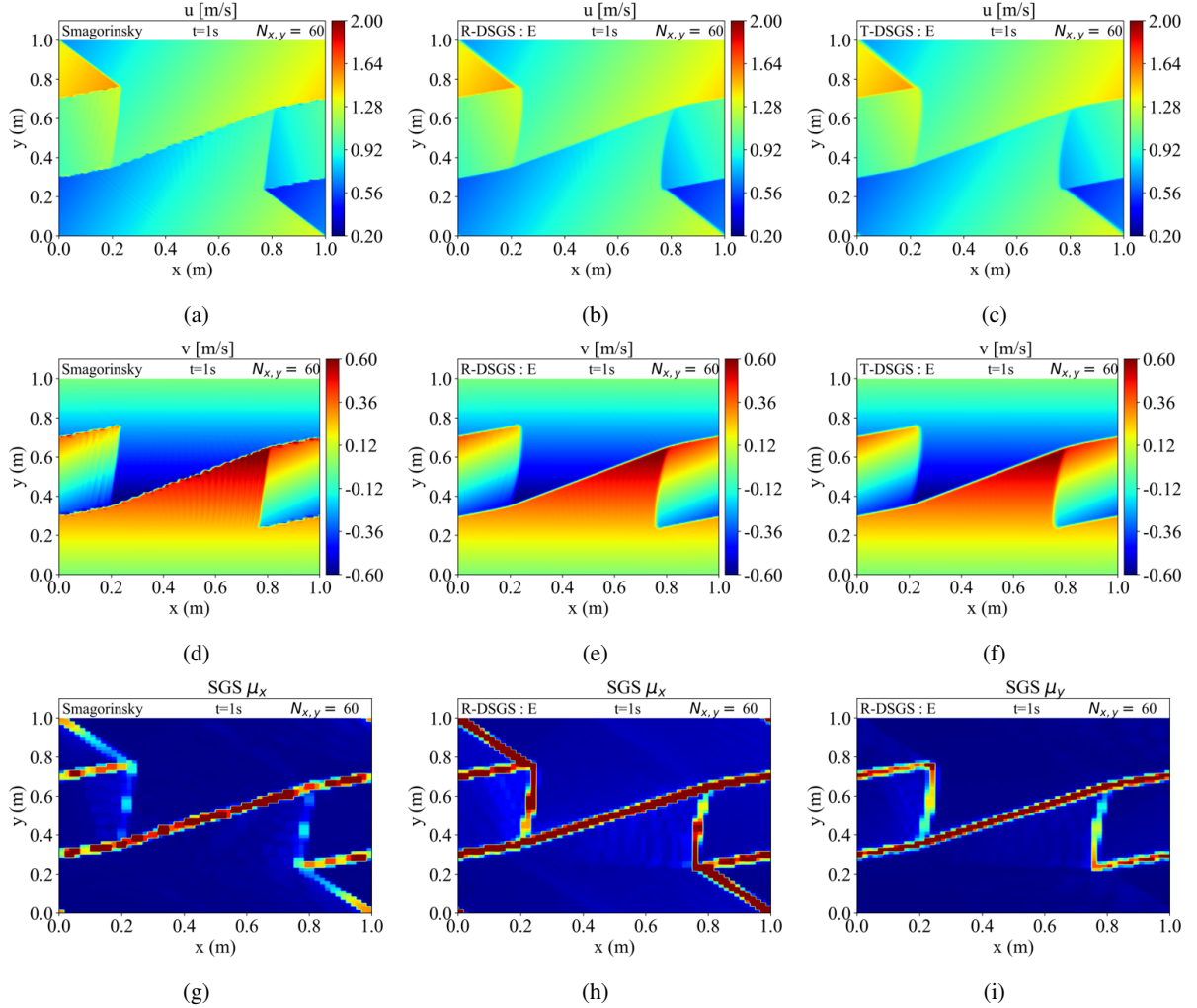

Figure 5: Solution of the 2D Burgers' equation at time $t = 1.0\ s$ showing $x$ velocity using a) Smagorinsky, b) R-DSGS and c) T-DSGS; $y$ velocity using d) Smagorinsky, e) R-DSGS and f) T-DSGS; and distribution of g) $\mu_{x,y}$ for Smagorinsky, h) $\mu_x$ for R-DSGS and i) $\mu_y$ for the R-DSGS (distribution of $\mu_{x,y}$ for the T-DSGS is similar to that of R-DSGS).

Figure 6 shows the solution to the Burgers' equation at $t = 1$ s along the line $y = 0.4$ m. The distribution of the solution obtained using the Vreman model is not presented as the solution diverged at $t = 0.7$ s. Although the Smagorinsky model yields the least dissipated solution, it also exhibits Gibbs' phenomena in both the $x$-velocity and $y$-velocity distribution. The elemental T-DSGS model yields the most dissipated solution, while the nodal R-DSGS yields the least dissipated, non-oscillatory solution. It should be noted that none of the DSGS models exhibit oscillations near the discontinuities and, therefore, are better able to stabilize the solution. The results for Burgers' equation shows that the DSGS models are the most robust and stable for nonlinear problems with discontinuities.

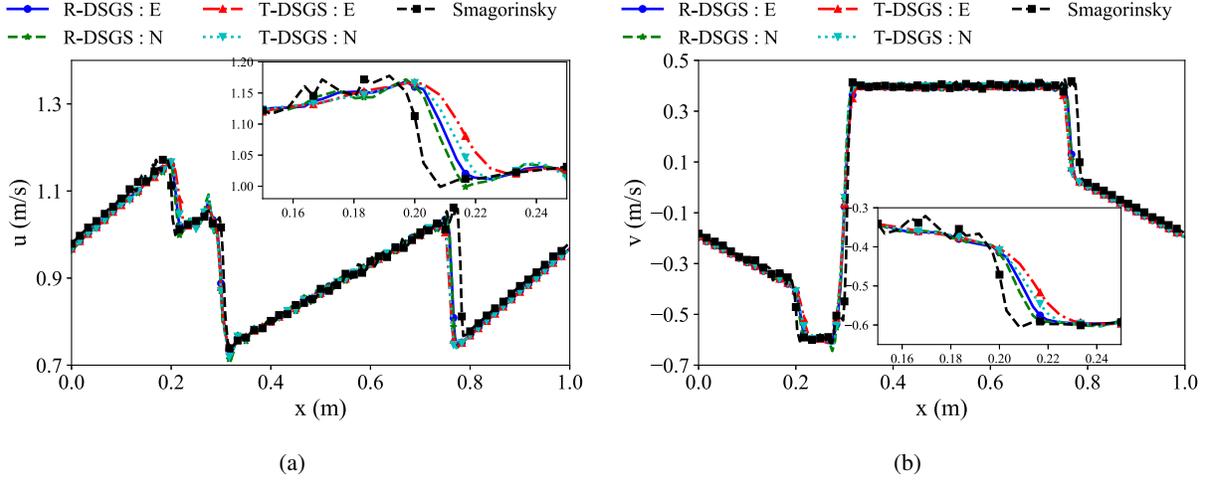

Figure 6: Solution of the 2D Burgers' equation at time $t = 1.0$ s along the line $y = 0.4$ m showing: a) the x-velocity and b) y-velocity

# 7 Case 3: Taylor-Green Vortex

## 7.1 Problem Definition

The Taylor-Green vortex (TGV) has been a canonical problem in computational fluid dynamics to benchmark algorithms for modeling vorticity, turbulence, and the decay of turbulence. First presented by G.I. Taylor and A.E. Green in 1935 [65], the TGV problem has been extensively used to analyze various SGS models [21, 66–70] and are often compared to the DNS results of Brachet et al. [71]. The TGV problem is initialized with a uniform density of $\rho = 1.178$ kg/m$^3$ and the velocity field given by

$$
\begin{aligned}
u(x, y, z) &= u_0 \sin(kx) \cos(ky) \cos(kz) \\
v(x, y, z) &= -u_0 \cos(kx) \sin(ky) \cos(kz) \\
w(x, y, z) &= 0
\end{aligned}
\tag{27}
$$

where $u_0 = 100$ m/s and $k$ is the wavenumber with a value of $k = 1$ m$^{-1}$. The pressure field is given by

$$
p = p_0 + \rho u_0^2 \left[ \frac{1}{16} \left( \cos(2kx) + \cos(2ky) \right) \left( \cos(2kz) + 2 \right) \right].
\tag{28}
$$

The flow domain is taken to be a triply-periodic (periodic in $x$, $y$ and $z$) cube of length $2\pi$ [21]. The time integration is performed using an explicit third-order five-stage Runge-Kutta method. We use an effective grid resolution $\Delta x = \Delta y = \Delta z \approx 0.049$ m with 32 elements in the $x$, $y$ and $z$ directions with fourth order polynomials.

The dynamics of the flow evolution are best analyzed by observing certain integral quantities. The volume-averaged kinetic energy is given by

$$
E = \frac{1}{\mathcal{V}} \int_{\mathcal{V}} \frac{1}{2} \left( \sum_{i=1}^{3} u_i u_i \right) d\mathcal{V} = \frac{1}{2} \left\langle |\mathbf{u}|^2 \right\rangle
\tag{29}
$$

where $\mathcal{V}$ is the volume of the domain and $\langle . \rangle$ is the volume averaging operator. We define $E^* = E/E_0$ as the normalized volume-averaged kinetic energy, where $E_0 = E(t = 0)$ is the quantity at time $t = 0$ s. The kinetic energy, in theory, should be conserved for inviscid dynamics and when the numerical scheme is able to resolve all scales of motion. Therefore, the decay of kinetic energy can be used to identify at which point the flow becomes under-resolved. Any change over time on the volume averaged kinetic energy is entirely[2] due to a term related to the SGS stress tensor known as the kinetic energy dissipation rate $\varepsilon$:

$$
-\frac{dE}{dt} = \varepsilon = \left\langle u_i \frac{\partial \tau_{ji}^{SGS}}{\partial x_j} \right\rangle.
\tag{30}
$$

---

[2]In low-mach flows the kinetic energy dissipation due to the pressure dilatation term $\left\langle p \frac{\partial u_i}{\partial x_i} \right\rangle$ is negligible.

The enstrophy, defined as $\langle \omega^2 \rangle = \left\langle |\nabla \times \mathbf{u}|^2 \right\rangle$, is an indicator of vorticity production. The normalized enstrophy $\langle \omega^2 \rangle^*$ is defined as $\langle \omega^2 \rangle^* = \langle \omega^2 \rangle / \langle \omega^2 \rangle_0$, where $\langle \omega^2 \rangle_0$ is the enstrophy at time $t = 0$ s. The enstrophy is related to the kinetic energy dissipation rate by $\varepsilon = \nu_{eff} \langle \omega^2 \rangle$ [72], where $\nu_{eff}$ is the effective viscosity. All results for the TGV test case are presented and analyzed in non-dimensional time $t^* = k u_0 t$.

## 7.2 Results

The effects of the SGS dissipation on the volume-averaged kinetic energy, kinetic energy dissipation rate, enstrophy and effective viscosity are presented. Figure 7 shows these four properties as a function of non-dimensional time $t^*$. All four quantities suggest that the elemental T-DSGS and R-DSGS models are the most dissipative of the six SGS models. It should be mentioned that the $C_s$ constant in the Smagorinsky model (Eq. (8)) is typically tuned for particular applications. In the same way, the DSGS models can also be tuned by selecting an appropriate $C$ parameter in Eq. (12). Figure 7a shows that the nodal R-DSGS model is better able to conserve the kinetic energy early on in the simulation ($t^* < 12$) but leads to more dissipated energy by $t^* \approx 24$. The elemental T-DSGS and R-DSGS models are more dissipative than the remaining four models until $t^* \approx 24$. At earlier times, the Vreman and Smagorinsky models are almost indistinguishable but the Vreman model overall conserves more kinetic energy as the solution progresses. Between $t^* \approx 28$ and $t^* \approx 40$, the Smagorinsky model results in more dissipated energy than the Vreman model.

The kinetic energy dissipation rate shown in Fig. 7b increases rapidly until $t^* \approx 9$, after which it decreases exponentially. A DNS study performed by Brachet et al. [71, 73] predicted that the maximum dissipation rate is reached at $t^* \approx 9$. This corresponds to a Reynolds number ($Re = u_0/k\nu$) of $Re > 3000$, which is similar to the Reynolds number encountered in this study. All of our SGS models experience this peak in dissipation rate at $t^* \approx 8$ and are in good agreement with the DNS results of Brachet et al. [71] and the LES results of Rafei et al. [21]. The R-DSGS models experience the highest peak in kinetic energy dissipation rate. The dissipation rate of the R-DSGS models is consistently higher than the remaining models thereby leading to more overall kinetic energy dissipation by $t^* \approx 24$. This is indicative of the energy being dissipated at smaller length scales. All models show a slight inflection at $t^* \approx 5$ which can indicate the breakdown of the vortex tubes into small-scale turbulence. Before this time, energy should be conserved since the smaller eddies are not present to dissipate the energy. This indicates the superior performance of nodal R-DSGS models as it is able to conserve more energy, before the breakdown of the vortex tubes, than the other SGS models.

Figure 7c shows the enstrophy evolution over time. For inviscid flow, the enstrophy should increase to infinity. The dissipation of the SGS models bounds this increase in vorticity. The peaks in enstrophy coincide approximately with the peaks in kinetic energy dissipation rate, and therefore, can also be used to determine when the flow becomes under-resolved. The Smagorinsky, Vreman and elemental DSGS models produce the lowest volume-averaged enstrophy. The nodal DSGS models feature the largest volume-averaged enstrophy. The magnitudes and distributions of the enstrophy are in good agreement with those obtained by Shu et al. [74], Drikakis et al. [69] and Rafei et al. [21] for nearly incompressible simulation of the Taylor-Green vortex.

Figure 7d shows the effective viscosity, which includes the dissipation due to the SGS model and any numerical dissipation present in the discretization scheme. The effective viscosities for all models experience an inflection at $t^* \approx 5$, coinciding with the inflection point which signals the breakdown of vortex tubes into small-scale turbulence. All models also experience a peak at $t^* \approx 8$ which coincides with the peak in dissipation rate. As expected, the elemental DSGS models show a higher effective viscosity while the nodal variants feature a lower, decreasing viscosity.

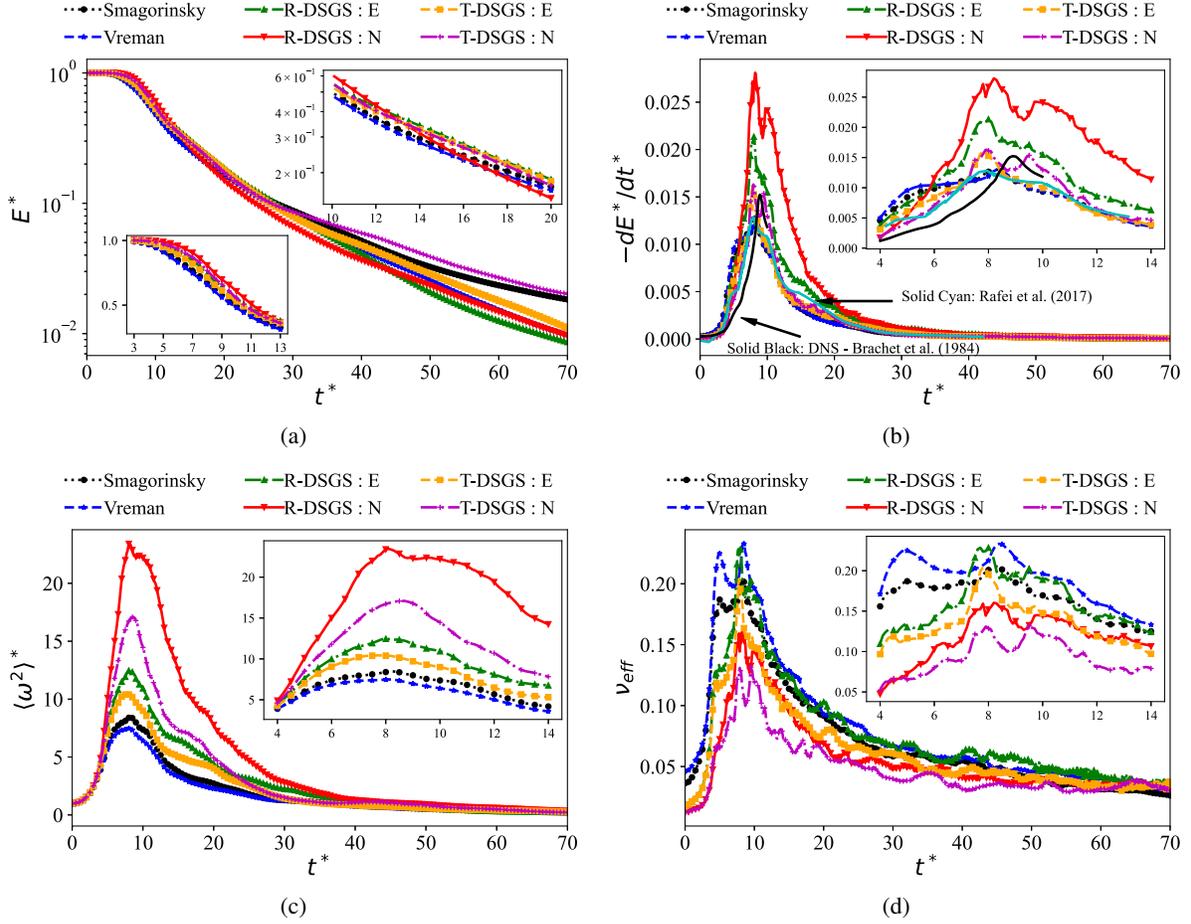

Figure 7: Evolution with non-dimensional time $t^*$ (with zoomed-in views) of: a) volume-averaged kinetic energy, b) kinetic energy dissipation rate, c) enstrophy and d) effective viscosity.

The kinetic energy spectra provide insight into the flow topology and energy cascade between different length scales. Figure 8 shows the kinetic energy spectra for the six SGS models. All models experience a peak at wavenumber $k \approx 3$. This peak indicates the imprint of the initial conditions at $t^* = 0$ still remaining in the TGV flow structure [69]. The spectrum with constant slope of $-5/3$ is referred to as the Kolmogorov spectrum and represents the slope of the energy spectrum, in homogenous isotropic turbulence, between the integral length scales and the Kolmogorov length scales. This region between the two length scales is known as the inertial subrange [75]. All models exhibit this $k^{-5/3}$ spectrum slope for $k < 32$.

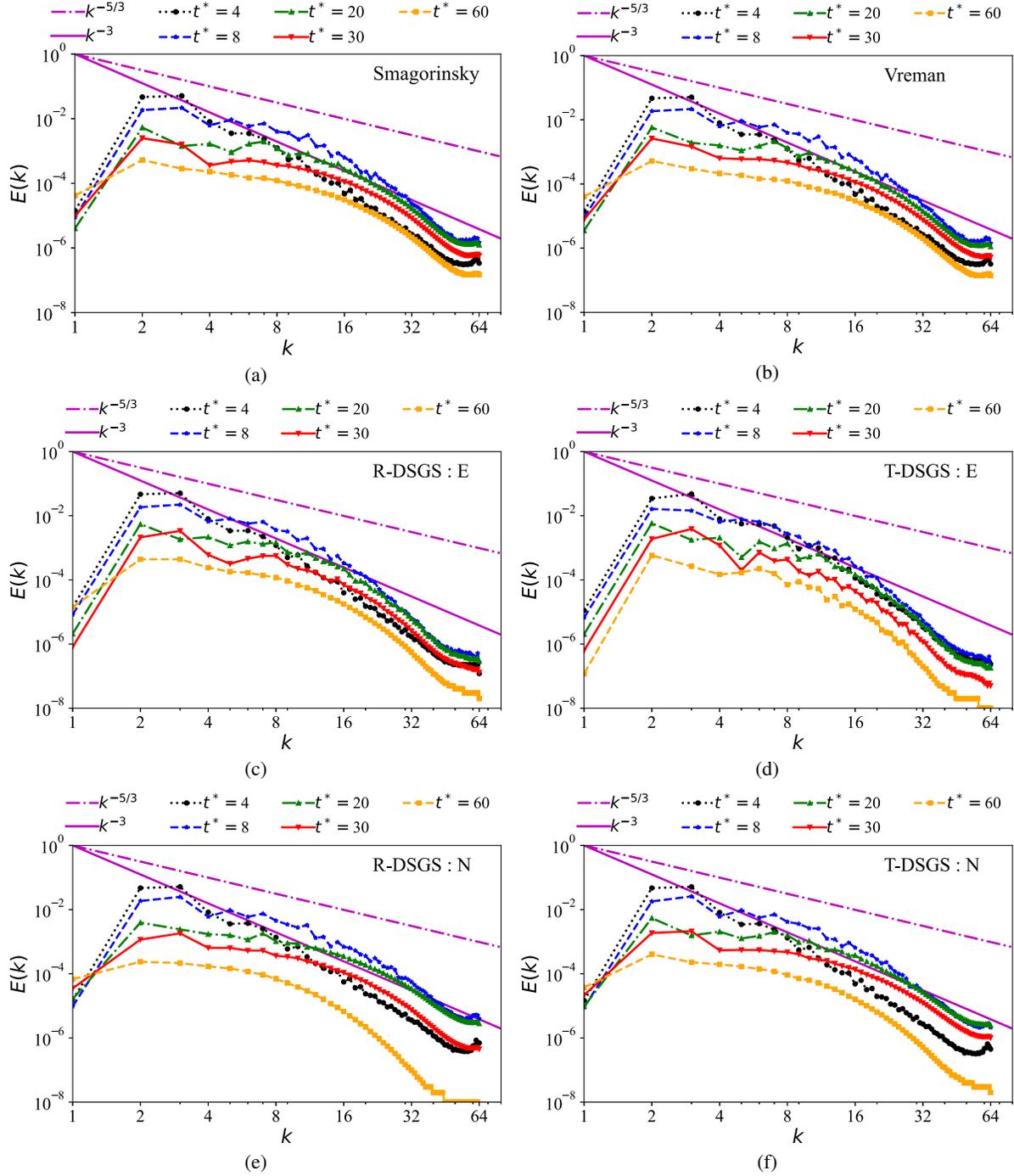

Figure 8: Kinetic energy spectra obtained using: a) Smagorinsky, b) Vreman, c) elemental R-DSGS, d) elemental T-DSGS, e) nodal R-DSGS and f) nodal T-DSGS.

## 8 Case 4: Thermal Density Current

### 8.1 Problem Definition

The 2D density current test problem, introduced in [76], is a standard benchmark for development and verification of atmospheric models [77–79]. It features a cold bubble of air descending to the ground. The density current then develops Kelvin-Helmholtz shear instability rotors as it spreads out laterally. This case is often solved using constant and uniform diffusion coefficient $\mu = 75$ m$^2$s$^{-1}$ [45, 76]. To our knowledge, only two works have analyzed it within the LES framework [39, 78]. This case investigates the ability of the SGS models to resolve the finer structures of these rotors.

The initial distribution of the cold air bubble is given as

$$\Delta\theta = \frac{\theta_c}{2}\left[1 + \cos\left(\pi_c r\right)\right] \quad \text{if } r \leq 1 \tag{31}$$

where $\theta_c = -15$ K and

$$r = \sqrt{\left(\frac{x - x_c}{x_r}\right)^2 + \left(\frac{z - z_c}{z_r}\right)^2}.$$

The domain is defined as $(x, z) \in [0, 25] \times [0, 6.4]$ km$^2$ with $t \in [0, 900]$ s and the center of the bubble with a size $(x_r, z_r) = (4, 2)$ km is positioned at $(x_c, z_c) = (0, 3)$ km. A no-flux boundary condition is imposed on all four boundaries of the domain. The three different effective grid resolutions $\Delta x = \Delta z = 50$ m, $\Delta x = \Delta z = 25$ m and $\Delta x = \Delta z = 12.5$ m with polynomial of order four are considered. The time integration is performed using an IMplicit-EXplicit (IMEX) method [47] which treats the fast moving acoustic and gravity waves implicitly and the advection terms explicitly. The IMEX method in this work is based on a second-order additive Runge-Kutta method (ARK). Since the continuity, momentum and energy equations are all coupled, their dynamics are influenced by the complete system, an additional passive tracer is also incorporated to investigate the effects of the SGS model on only the transport equations [3]. The tracer concentration follows the same distribution as $\Delta\theta$ using $q_c = 15$ kg/kg. In this case, a single diffusion coefficient is computed by the DSGS models for both ($x$ and $y$) momentum balance equations so that the same eddy viscosity is applied to both momentum equations, as is done in the other SGS models.

## 8.2 Results

Figures 10 shows the potential temperature distribution at $t = 900$ s using the six SGS models at three different resolutions. In all cases, increasing the grid resolution yields more resolved structures.

The Kelvin-Helmholtz (KH) instability rotors only become visible at a grid resolution of $\Delta x = 50$ m when using Smagorinsky and Vreman models. These rotors become visible for the DSGS models at resolution of $\Delta x = 25$ m and at $\Delta x = 12.5$ m. Comparing Figs. 10i and 10n, it can be seen that the nodal DSGS models are able to resolve the rotor structure at a lower grid resolution than the elemental DSGS models. A similar conclusion can be drawn by comparing Figs. 10l and 10n. Since the potential temperature ($\theta$) is coupled to the mass and momentum balance equation through the equation-of-state, its extrema are not bound by the initial distribution. On the other hand, since the tracer is passive and transported by the velocity field, its minima are bound. The figures showing the tracer distribution are provided as Supplementary Material. They show that the Smagorinsky and Vreman models exhibit the largest overshoots and undershoots (as seen by the negative tracer concentration). The elemental DSGS models damp these undershoots.

Figures 9a - Fig. 9c show the evolution of enstrophy and the kinetic energy spectrum at $t = 900$ s. As expected, the enstrophy generated increases as the mesh is refined and more circulating rotors are resolved. The Smagorinsky and Vreman models result in similar enstrophy values and are consistently higher than the other SGS models. The enstrophy generation for all SGS models (and constant diffusion) follows a similar trend until $t = 300$ s, which coincides with the point at which the flow spreads laterally and begins to develop the first swirling vortex. At finer resolutions, the enstrophy in the flow continues to increase indicating the increased shedding of Kelvin-Helmholtz rotors. We note that the DSGS models can also be tuned by selecting an appropriate parameter $C$ in Eq. (12). The effects of this parameter $C$ on the DSGS models will be studied in future work.

Figures 9d - Fig. 9f show the kinetic energy spectra of the flow at $t = 900$ s. The spectra were computed using data in the window $x \in [5000, 15000]$ m and $z \in [0, 4000]$ m. The data set was made periodic using the approach of Errico [80]. This approach makes the data periodic by removing the larger linear trends and was used by Skamarock [81] to compute the spectra in atmospheric flows. It can be seen that for lower wavenumbers $k$, all SGS models follow Kolmogorov's $k^{-5/3}$ spectrum slope. For higher wavenumbers, all SGS models achieve Kraichnan's $k^{-3}$ spectrum slope [82], which is expected in 2D turbulent flow. Similar spectra were observed for other atmospheric flows [81, 83].

---

[3]The figures showing the tracer distribution are provided as Supplementary Material.

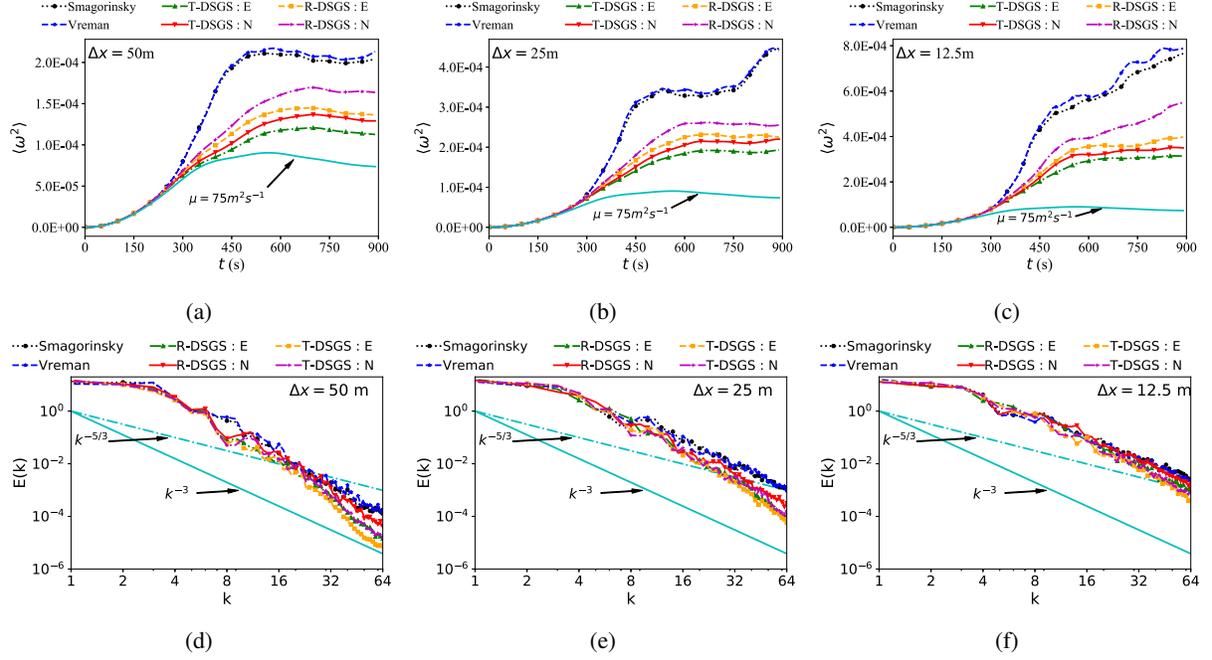

Figure 9: Evolution with time of the turbulent energy dissipation rate at a resolution of: a) 50 m, b) 25 m, c) 12.5 m; and kinetic energy spectrum at a resolution of d) 50 m, e) 25 m, f) 12.5 m

Table 2 shows the location of the front obtained using various SGS models and discretization schemes. It includes models used in this work and those reported in the open literature. The discretization schemes considered in the table include the discontinuous Galerkin (DG), spectral element (SEM), finite element (FEM), finite volume (FVM), and finite difference (FDM) methods. Out of those taken from the literature, only the variational multi-scale (VMS) model uses a non-constant diffusion coefficient. The remaining tabulated models assume a constant viscosity of $\mu = 75$ m$^2$s$^{-1}$. For more information regarding these models, the reader is referred to the references. Table 2 shows that the front moves further ahead with increasing grid resolution, indicating a faster moving flow. This is seen for all SGS models except the Smagorinsky SGS model. These lower speeds at lower resolutions can be an indication of larger dissipation. The diffusion coefficient computed by each of the SGS models is a quadratic function of the filter-width (element size). Therefore, increasing the element size (i.e. decreasing the resolution) yields larger dissipation. The front location can be a qualitative indicator of the magnitude of the dissipation of the SGS model.

For the R-DSGS model, this conclusion is not as straightforward. Although decreasing the resolution increases the filter-width (and the dissipation), it can also reduce the oscillations in the solution and, therefore, the residual. In each case studied in this work, the front locations at a resolution of $\Delta x = 50$ m lie within 14600 m and 15040 m and are in good agreement with the bounds reported in the literature (within 14400 and 15030 m). The front locations at resolutions of $\Delta x = 25$ m lie within 14900 m and 15120 m, and the front locations at resolutions of $\Delta x = 12.5$ m lie within 15075 m and 15160 m. This test case shows that the DSGS models exhibit the expected traits of 2D turbulence and that the R-DSGS models produce the least diffused rotor structures. The results from all the SGS models are also in agreement with those presented in the open literature.

Table 2: Front location (m) at $t = 900$ s using various SGS models at different resolutions.

| Model | Discretization | SGS Model | $\Delta x = 50$ m | $\Delta x = 25$ m | $\Delta x = 12.5$ m |
|---|---|---|---|---|---|
| NUMA (Current) | SEM | Smagorinsky | 15034 | 15117 | 15091 |
| | SEM | Vreman | 14834 | 15082 | 15125 |
| | SEM | R-DSGS : E | 14634 | 14950 | 15158 |
| | SEM | T-DSGS : E | 14600 | 14900 | 15075 |
| | SEM | R-DSGS : N | 14634 | 14917 | 15125 |
| | SEM | T-DSGS : N | 14634 | 14950 | 15108 |
| NUMA [39] | SEM | Smagorinsky | 14726 | 14918 | - |
| | SEM | Dyn-SGS | 14535 | 14992 | 15056 |
| ALYA [84] | FEM | VMS [30, 85] | 14629 | 14890 | - |
| Cases with constant viscosity: $\mu = 75$ m$^2$s$^{-1}$ | | | | | |
| NUMA [45] | SEM | $\mu = 75$ m$^2$s$^{-1}$ | 14629 | - | - |
| | DG | | 14767 | - | - |
| $f$-wave [86] | FVM | $\mu = 75$ m$^2$s$^{-1}$ | 14975 | - | - |
| WRF-ARW [86] | FDM | | 14470 | - | - |
| REFC [76] | FDM | | 14437 | - | - |
| REFQ [76] | FDM | | 14409 | - | - |
| PPM [76] | FDM | | 15027 | - | - |

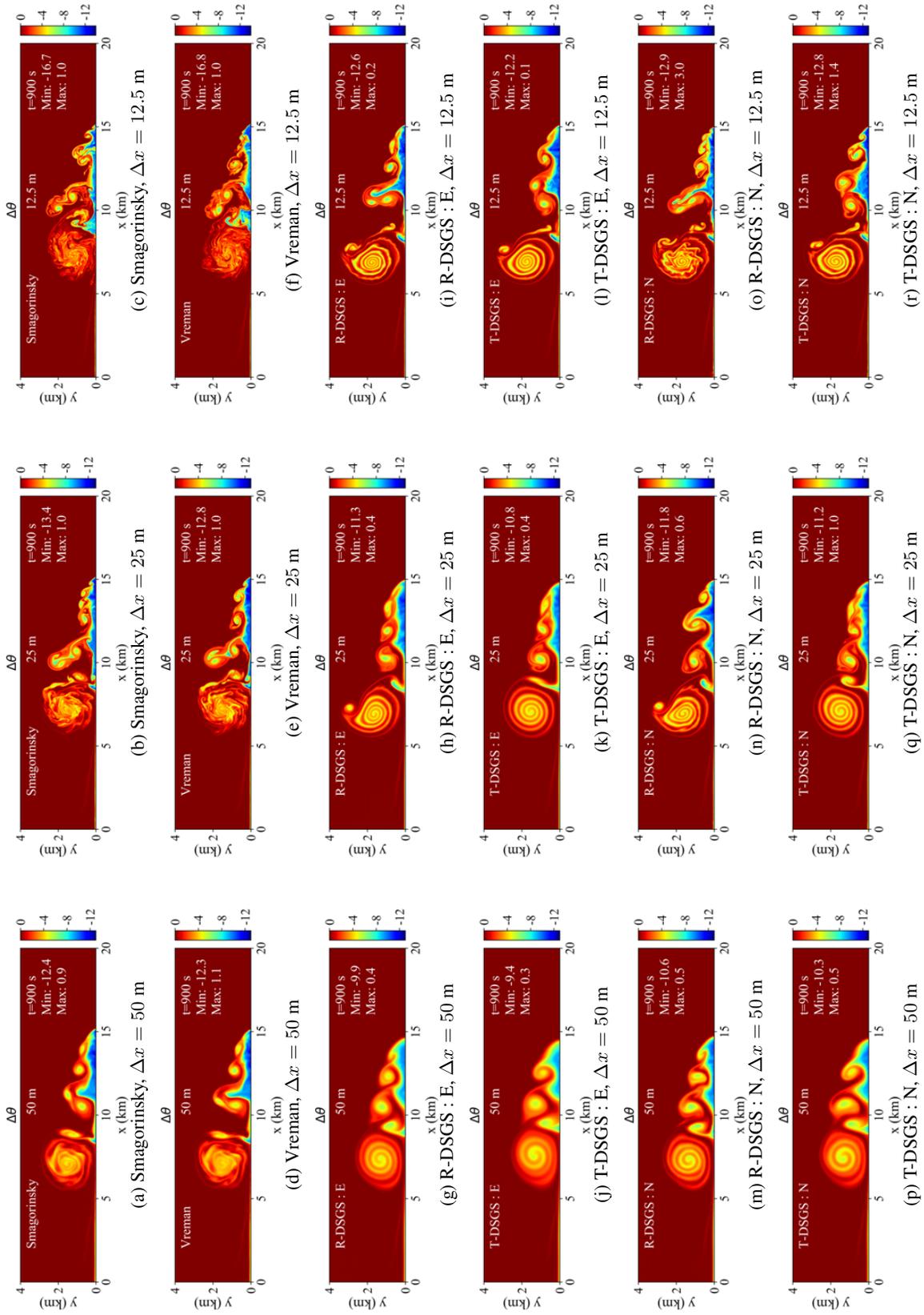

Figure 10: Distribution of potential temperature perturbation ($\Delta\theta$) for 2D density current problem at $t = 900$ s for various SGS models at different resolutions

# 9    Conclusions

In this study we perform a comparison of various SGS models on various test problems. Two common SGS models (Smagorinsky and Vreman) are compared to two newly developed models, R-DSGS and T-DSGS. The SGS models' superiority are judged based on their ability to capture discontinuities, dampen overshoots and undershoots, stabilize the solution, and resolve the flow structure. Although it is common to consider the turbulent diffusion coefficient to be constant within the element and piecewise constant over the domain, this work also considers a higher order representation of the coefficients. The SGS models are evaluated for problems with passive tracer advection in a deforming flow field, discontinuities (Burgers' equation), homogeneous isotropic turbulence (Taylor-Green Vortex), and thermally driven flows (density current).

The analysis using a discontinuous distribution of a passive tracer in a deforming flow field (Case 1) shows that the Smagorinsky, Vreman, and nodal R-DSGS models are all able to preserve the discontinuous interface. The Smagorinsky model is shown to yield a sharp, more resolved discontinuous interface but also exhibited Gibbs phenomenon (oscillations) whereas the Vreman and nodal R-DSGS models are able to damp these oscillations while preserving the sharp interface. The more dissipative T-DSGS models (elemental and nodal) damp these Gibbs' oscillations but also diffuse the interface. In general, the elemental DSGS models are more diffusive than the nodal DSGS models due to the larger diffusion coefficients arising from the infinity-norms. The T-DSGS models are also more diffusive than the R-DSGS models since the unsteadiness of the flow is typically much larger than the residual of the PDE.

The 2D, two-equation Burgers' system (Case 2) is used to evaluate the SGS models' ability to not only preserve and resolve the discontinuities but also to stabilize the solution. The R-DSGS and T-DSGS models are able to both preserve the sharp interface and stabilize the solution. The Gibbs' oscillations arise in both the Smagorinsky and Vreman models. These oscillations continue to grow thereby driving the solution obtained using the Vreman model unstable at $t = 0.7\ s$. Although the Smagorinsky model is able to preserve stability until $t = 0.7\ s$, it did result in severe oscillations near the discontinuities. Whereas the Smagorinsky and Vreman models stabilize the coupled system using a single diffusion coefficient, the DSGS models decouple the stabilization of the two-equation Burgers' system, thereby stabilizing each equation individually. Therefore, the DSGS models are able to continue well past the point of failure of the Smagorinsky and Vreman models. The DSGS models do not show such oscillations and, therefore, are the most stable SGS models studied.

The Taylor-Green Vortex problem (Case 3) is used to investigate the SGS models' ability to transfer energy from larger scales to smaller scales. At early simulation times, before the breakdown of the vortex tubes, the nodal R-DSGS model is shown to be the least dissipative and most energy conserving. This is a strong indicator of the nodal R-DSGS model's superior performance since energy before the vortex breakdown should be conserved as there are no small-scale eddies present to dissipate it. The kinetic energy dissipation rates for all models are in good agreement with the results predicted by DNS.

The thermally driven density current (Case 4) problem is used to evaluate the models' ability to resolve finer flow structures. As expected, the flow structures become more resolved with increasing grid resolution. The Kelvin-Helmholtz instability rotors are visible at a coarser resolution when using the Smagorinsky and Vreman models. The rotor structures obtained using element DSGS models at finer resolutions are seen at much coarser resolutions when using their nodal variants. The enstrophy evolution indicates that the Smagorinsky and Vreman models are the least dissipative. All SGS models are shown to exhibit the Kolmogorov-Kraichnan spectra expected in 2D turbulence.

In summary, the following conclusion can be drawn for the SGS models compared in this study:

1. Smagorinsky, Vreman, and R-DSGS are able to capture and preserve sharp discontinuities.

2. Smagorinsky and Vreman exhibit Gibbs' phenomenon near discontinuities. This is damped by the DSGS models.

3. DSGS models are more robust than Vreman and Smagorinsky when utilized for stabilizing purposes.

4. The Smagorinsky and Vreman models exhibit larger overshoots and undershoots, but are able to resolve the Kelvin-Helmholtz rotor structures at much lower resolutions.

5. Nodal R-DSGS is better able to preserve the kinetic energy.

6. All SGS models exhibit the Kolmogorov-Kraichnan spectrum.

The different analyses show that the nodal R-DSGS model is able to accurately capture and preserve the discontinuities (Case 1 and 2), stabilize the solution (Case 2), conserve more kinetic energy and then dissipate it to smaller length scales (Case 3), and resolve finer flow structures (Case 4).

## Acknowledgments

The NPS authors are grateful for the research funding provided by the Office of Naval Research under grant # N0001419WX00721. This research was performed while S. Reddy and F.A.V. de Bragança Alves held NRC Research Associateship awards at the Naval Postgraduate School. The authors are also grateful for Stephen Guimond for providing his source code for computing the kinetic energy spectra.